# A Two-Stage AI-Powered Motif Mining Method for Efficient Power System Topological Analysis

Yiyan Li [a,c], Zhenghao Zhou [a,c], Jian Ping [a,c], Xiaoyuan Xu [b,c], Zheng Yan [b,c*], Jianzhong Wu [d]

[a] *College of Smart Energy, Shanghai Jiao Tong University, Shanghai, 200240, China*
[b] *The Key Laboratory of Control of Power Transmission and Conversion, Ministry of Education, Shanghai Jiao Tong University, Shanghai 200240, China*
[c] *Shanghai Non-Carbon Energy Conversion and Utilization Institute, Shanghai Jiao Tong University, Shanghai 200240, China*
[d] *School of Engineering, Cardiff University, Cardiff, UK*

**Abstract:** Graph motif, defined as the microstructure that appears repeatedly in a large graph, reveals important topological characteristics of the large graph and has gained increasing attention in power system analysis regarding reliability, vulnerability and resiliency. However, searching motifs within the large-scale power system is extremely computationally challenging and even infeasible, which undermines the value of motif analysis in practice. In this paper, we introduce a two-stage AI-powered motif mining method to enable efficient and wide-range motif analysis in power systems. In the first stage, a representation learning method with specially designed network structure and loss function is proposed to achieve ordered embedding for the power system topology, simplifying the subgraph isomorphic problem into a vector comparison problem. In the second stage, under the guidance of the ordered embedding space, a greedy-search-based motif growing algorithm is introduced to quickly obtain the motifs without traversal searching. A case study based on a power system database including 61 circuit models demonstrates the effectiveness of the proposed method.

## 1. Introduction

After decades of development, the power grid has evolved into a hyper-complex physical system. The large-scale integration of renewable energy resources and flexible loads accelerates the electronization of the power system, leading to low system inertia and high operation uncertainty. Meanwhile, advanced operation modes such as peer-to-peer energy trading, demand response and virtual power plants put further challenges to the system operation. Therefore, it is important to continually deepen the understanding of the fast-developing power grid topology for both planning and operation purposes.

Because the power grid naturally has a graph structure with nodes and edges, complex network theory has been widely used to analyze the power system from the topological perspective. In [1], the North American power grid is analyzed from a topological perspective to determine the power transfer capability and identify key transmission substations that make the grid vulnerable when being disrupted. Similar study is conducted by [2], using complex network theory to analyze the static robustness of the European power grid when failures or attacks happen. Ref [3] implements graph neural network to achieve fast reliability evaluation for the park-level electricity-hydrogen system by analyzing the topological structures. A feature-selection algorithm is introduced to assess the importance of the system components to the evaluation results. Focusing on bulk power systems, [4] proposes an efficient topological procedure to calculate the reliability indices and determine the contribution of each system state to those indices. Ref [5] focuses instead on the medium and low voltage grids to analyze the adequacy of the current distribution infrastructure for a decentralized energy market, based on statistical tools from the complex network theory. Ref [6] examines the effect of network topology on the time-invariant dynamics of power systems. Stability metrics based on $H_2$-norm are defined to assist the planning for new or existing power systems. Note that purely relying on topological analysis may ignore the unique characteristics of power systems. To address this issue, [7] introduces an extended topological method to rank the most critical lines and buses by incorporating several physical features of power systems. Ref [8] considers electrical features of power systems such as electrical distance, power transfer distribution and line flow limits. An extended betweenness centrality is then defined to identify critical components in power grids.

As one of the major characteristics of graphs, *topology motif* has gained increasing attention in recent years in analyzing power system stability, vulnerability and resiliency. As shown in Fig. 1, motif is defined as the microstructure of a complex network that appears more frequently than other microstructures, which is considered as the basic building block of a complex network and is related with many global topological characteristics [9]. Ref [10] proposes a motif-based power grid robustness analysis method, focusing on the influence of local network structures to the system robustness under attacks. Ref [11] demonstrates that the grid fragility tends to increase when the grid becomes more interconnected and complex motifs starts to appear, based on the assessment of European grid reliability. Based on the topological analysis of power grids, Ref [12] concludes that certain motif structures such as "dead trees" will degrade the overall dynamic stability of the system. On the contrary, the appearance of certain motifs in the power grid such as three-cycles is helpful to the dynamical stability of the power grid [13]. Ref [14] studies the reliability of a cyber-physical system based on motif analysis, considering the influence of both the cyber layer and the physical layer. Ref [15] implements motif theory to analyze the multiple line outage patterns in the transmission systems. Results show that multiple line outages usually have certain spatial patterns that occur much more frequently than random selection, which are defined as contingency motifs. The *N-k* contingency lists constructed from motifs can improve risk estimation in cascading outage simulations and help to confirm utility contingency selection.

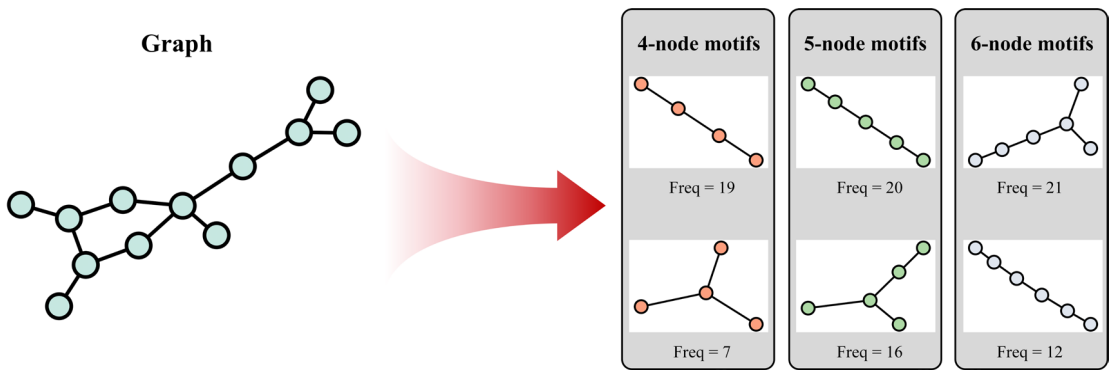

Fig. 1 Illustration of the graph motif concept. Frequency is calculated by traversal searching.

However, as the value of motif is being recognized in power system analysis, how to efficiently find motifs in large-scale power systems is extremely computationally challenging because of the following three reasons [16]. *First*, according to the motif definition, the key to find motifs is to count the appearances of the candidate microstructures in the target large graph, which is a typical NP-hard problem [17]. *Second*, there exists the graph isomorphic issue when matching the motif candidates to the large graph, which brings additional complexity to the searching process. *Third*, as the number of nodes increases, the possible structures of the candidate motifs increase exponentially. All the above reasons triple the complexity of the motif mining process, making the traditional traversal searching methods infeasible especially in large graphs such as power systems. As a result, the value of motif in power system analysis is still far from being fully exploited. For instance, ref [10][11][13][14] all focus on the simplest 4-node motifs to conduct analysis, which is considered limited without exploring more sophisticated structures to better reflect power system topological characteristics. To address this issue, artificial-intelligence-based (AI-based) approaches have been proposed in recent years in the computer science domain to accelerate the motif mining process [18]-[20], which brings new opportunity for the motif analysis in power system domain.

In this paper, we introduce a two-stage AI-powered motif mining method to enable efficient and wide-range motif analysis for large-scale power systems, inspired by the works of Rex Ying et al [21][22]. In the first stage, a representation learning method with specially designed network structure and loss function is proposed to encode power system topologies into vectors, forming the ordered embedding space. In the second stage, guided by the ordered embedding space, a greedy-search algorithm is proposed to efficiently grow motifs in the target system without traversal searching. The major contributions and novelties of the paper are as follows:

(1) We introduce an AI-powered motif mining method to achieve efficient and wide-range motif mining for large-scale power systems, which brings new opportunity for power system topological analysis. To the best of the authors' knowledge, this is also the first work of implementing AI to support the motif-based power system analysis.

(2) We propose a representation learning method to encode the power system topology into vectors whose spatial positions represent the sophisticated subgraph isomorphic relationship among graphs. This method significantly simplifies the graph matching process and can provide references for the application of interpretable machine learning in power systems.

(3) We propose a greedy-search-based motif growing algorithm to directly obtain the motifs in the target system, so that the enumeration costs on the exponentially-increasing motif candidates can be avoided.

The rest of the paper are organized as follows: Section II introduces the proposed AI-powered motif mining method, including ordered embedding and motif searching. Section III demonstrates the case study results based on a power system database. Section IV concludes this paper.

## 2. Methodology

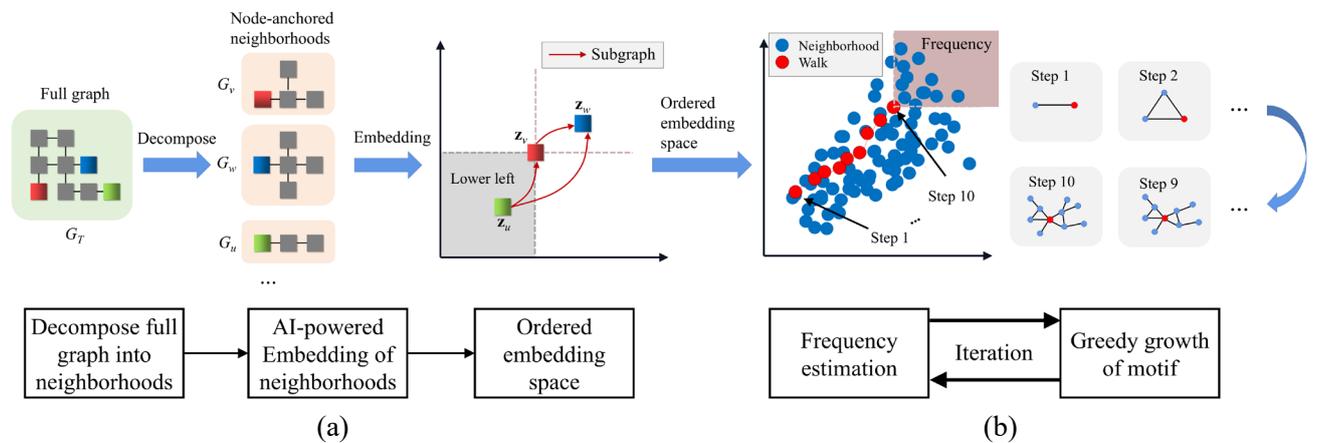

Fig. 2 Overview of the proposed motif mining process. (a) The ordered embedding process to encode graphs into vectors to formulate the ordered embedding space. (b) The motif searching process to iteratively grow the motifs.

As shown in Fig. 2, the proposed motif mining method includes two stages: *ordered embedding* and *motif*

*searching*. During the ordered embedding process, graphs are encoded as high dimensional vectors whose spatial positions represent their subgraph isomorphic relationships, formulating the ordered embedding space. Based on the ordered embedding space, the graph motifs can be obtained by the motif searching algorithm, which essentially is an iterative graph growing process instead of direct traversal searching so that the NP-hard problem can be avoided.

## 2.1 Basic Concepts of Graph

*Graph*: Graph-structured data $G$ is defined as $G=(V, E)$, where $V$ is the set of nodes of the graph, $E$ is the set of edges of the graph.

*Subgraph*: A graph $G'=(V', E')$ is called a subgraph of $G$ if $V'\in V$ and $E'=\{(u, v)\in E \mid u, v\in V'\}$, where $u$, $v$ are nodes.

*Graph isomorphism*: If there exists a one-to-one mapping between the nodes of graphs $G$ and another graph $H$, then $G$ and $H$ are defined as isomorphic. That is, there exists an edge between the two nodes of $G$ if and only if there is a corresponding edge between the two mapping nodes of $H$ [17].

*Subgraph isomorphism*: If a subgraph $G'\subseteq G$ is isomorphic to another graph $H$, then $H$ and $G$ are defined as subgraph isomorphism. The frequency of subgraph isomorphism is defined as the number of different $G'\subseteq G$ that are isomorphic to $H$. Two graphs are considered different if they do not share all nodes and edges.

*Graph motif*: Graph motif is defined as the small subgraphs of a large-scale graph that occur much more frequent than that of a random graph (i.e. frequent subgraph) [23]. Graph motifs usually include fundamental structural information of a complex graph and thus can be used to characterize the graph and derive new graphs.

*Neighborhood*: Neighborhood is defined as the subgraph anchored at a given node. A $k$-hop neighborhood anchored at node $v$ includes all nodes whose shortest path to $v$ is no more than $k$, along with the edge paths connecting these nodes.

## 2.2 Ordered Embedding – Theory

As shown in Fig. 2(a), ordered embedding space is a high dimensional vector space where each vector represents a graph. The unique property of this space is that the spatial locations of the vectors correspond to the subgraph isomorphic relationships among graphs. For example, if graph $G_u$ is a subgraph of $G_v$, then in the ordered embedding space the vector $\mathbf{z}_u$ (encoded from $G_u$) is on the "*lower left*" of the vector $\mathbf{z}_v$ (encoded from $G_v$)

$$\mathbf{z}_u[i]\leq \mathbf{z}_v[i] \ \forall_{i=1}^{D}, \ \ iff \ G_u \subseteq G_v \tag{1}$$

where $D$ is the vector dimension. Based on the ordered embedding space, the complex and time-consuming subgraph isomorphic problem is transformed into simple and efficient position comparing problem among vectors, which is the foundation of the proposed motif mining method.

To establish the ordered embedding space, inspired by [24], we design the following loss function to train a neural network encoder

$$L_{\mathbf{\theta}}(\mathbf{z}_u, \mathbf{z}_v) = \sum_{(\mathbf{z}_u, \mathbf{z}_v)\in \mathbf{P}} E(\mathbf{z}_u, \mathbf{z}_v) + \sum_{(\mathbf{z}_u, \mathbf{z}_v)\in \mathbf{N}} \max\{0, \alpha - E(\mathbf{z}_u, \mathbf{z}_v)\} \tag{2}$$

$$E(\mathbf{z}_u, \mathbf{z}_v) = \left\|\max\{0, \mathbf{z}_u - \mathbf{z}_v\}\right\|^2 \tag{3}$$

$\mathbf{P}$ represents the set of positive samples $(\mathbf{z}_u, \mathbf{z}_v)$ where the corresponding $G_u$ is a subgraph of $G_v$ (i.e. formula (1) is satisfied). $\mathbf{N}$ represents the set of negative samples where $G_u$ is a not subgraph of $G_v$ (i.e. formula (1) is violated). $\mathbf{\theta}$ is the parameter set of the neural network encoder. $\alpha$ is a positive hyper-parameter. $E$ is a function to measure the distance between two vectors. Equation (2) can be minimized if and only if all positive samples follow the "lower left" positional pattern in the embedding space while all negative samples violate the pattern. As a result, equation (2) will guide the neural network encoder to optimize its parameter $\boldsymbol{\theta}$ to achieve ordered embedding for graphs.

In practice, considering the neural network encoder can hardly be perfectly trained, we further define the binary subgraph prediction function $f(\mathbf{z}_u, \mathbf{z}_v)$ as follows to relax the subgraph matching criterion

$$f(\mathbf{z}_u, \mathbf{z}_v) = \begin{cases} 1 & iff \quad E(\mathbf{z}_u, \mathbf{z}_v) < t \\ 0 & otherwise \end{cases} \tag{4}$$

where $t$ is a positive value representing the threshold. $G_u$ is determined as a subgraph of $G_v$ if $f(\mathbf{z}_u, \mathbf{z}_v) = 1$. Note that according to equation (4), $G_u$ is still considered a subgraph of $G_v$ even if the "lower left" pattern is slightly violated within the given threshold $t$.

Based on the ordered embedding space and the subgraph prediction function in (4), we can quickly approximate the frequency of a given subgraph (i.e. the query graph $G_q$) that appears in the target graph $G_T$ to find out the most frequent subgraph (i.e. the motif). First, $G_T$ is randomly decomposed into multiple small neighborhoods (overlapping is allowed) through breadth-first searching [25]. Second, all neighborhoods will be encoded into the ordered embedding space by the trained encoder to formulate the reference vectors representing $G_T$. Meanwhile, $G_q$ will also be encoded into the ordered embedding space. Finally, the frequency of $G_q$ can be approximated by counting the number of reference vectors that satisfy equation (4). Note that if graph $G_u$ is a subgraph of $G_v$, then the frequency of $G_u$ is no less than $G_v$.

## 2.3 Ordered Embedding – Network Structure

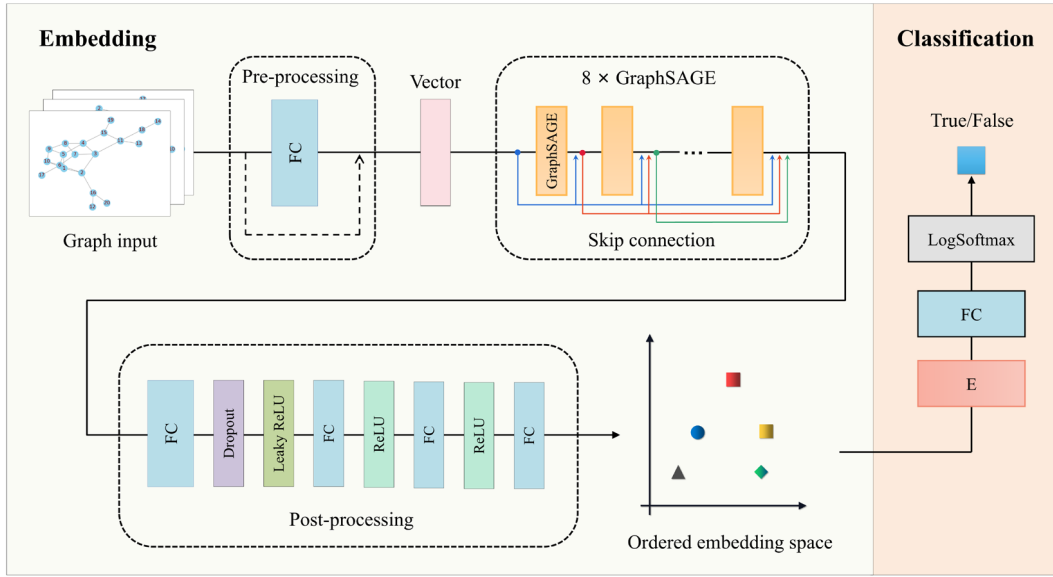

Fig. 3. Framework of the encoder network.

We design a two-stage neural network as the encoder to achieve the ordered embedding, shown as Fig. 3. The encoder network includes two stages: *embedding* and *classification*. In the embedding stage, the input graphs are converted to vectors by the pre-processing network, learned by the graph convolution network, and mapped into the ordered embedding space by the post-processing network. In the classification stage, the vectors in the ordered embedding space are fed into the subgraph prediction function in (4) to check whether they have subgraph isomorphic relationship.

When training the encoder, the model inputs include both positive sample set $\mathbf{P}$ and negative sample set $\mathbf{N}$, while the model outputs are binary labels. As a result, the encoder can be trained as a typical supervised learning problem. Note that once the encoder network is well-trained, it can be generalized across different domains. This is because the subgraph isomorphic patterns learned by the network is high-level topological knowledge that is independent of specific domains.

Specifically, the encoder network has the following distinct structural characteristics.

### GraphSAGE

Graph Convolutional Network (GCN) has been widely used as a machine learning method for graph-structured data [26]. However, as a transductive learning method, GCN captures the global features of the whole graph, which is computationally expensive when applied to the embedding problem of dynamic graphs. To address this issue, Graph Sample and Aggregation (GraphSAGE) is proposed in [27]. GraphSAGE is an inductive learning method that can characterize a given node by aggregating the information of its adjacent nodes, shown as equation (5).

$$\mathbf{h}_v^{(k)} = \sigma(\mathbf{W}^{(k)} \cdot f_k(\mathbf{h}_u^{(k-1)} : u \in \mathcal{N}(v))) \tag{5}$$

where $\mathbf{h}_v^{(k)}$ is the vector embedding of node $v$ at layer $k$, $\sigma$ is a non-linear activation function, $\mathbf{W}^{(k)}$ is the trainable weight matrix at lager $k$, $\mathcal{N}(v)$ is the set of the adjacent nodes of node $v$, $f_k$ is the aggregation function. In this paper we use the sum aggregation function to capture the cumulative influence of the adjacent nodes.

Fig. 4 shows the calculation process of a 2-layer GraphSAGE network as an example. In the 1st layer, each node aggregates its 1-hop adjacent nodes to obtain the vector embedding $\mathbf{h}^{(1)}$. In the 2nd layer, $\mathbf{h}^{(1)}$ on the 1-hop adjacent nodes of each node are further aggregated to $\mathbf{h}^{(2)}$. Such aggregation process continues until the final vector embedding is obtained. Note that by using a $k$-layer GraphSAGE network, we essentially capture the $k$-hop neighborhood $G_v$ centered at node $v$. Thus, the embedding of $v$ is equivalent to the embedding of $G_v$.

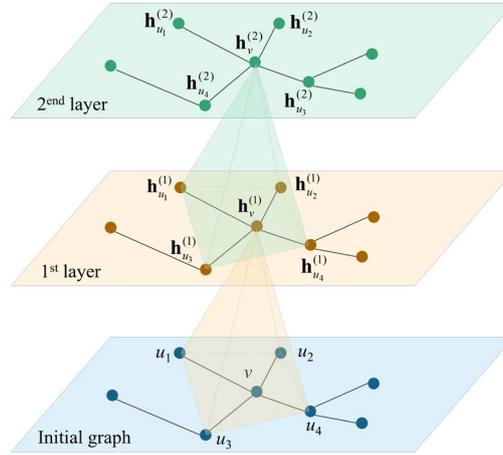

Fig. 4. Demonstration of GraphSAGE calculation process.

When training GraphSAGE, the number of the adjacent nodes $N$ ($N$=4 in Fig. 4) to be aggregated will first be specified as a fixed hyper-parameter. For each target node, if the number of its adjacent nodes is less than $N$, then sampling with replacement will be used until $N$ neighbors are reached. If greater than $N$, then random sampling without replacement will be used. After GraphSAGE is trained, we can easily obtain the embedding for a new node by aggregating its neighbors, which is more flexible and computationally efficient than GCN in dynamic graphs.

*Skip connections*

Based on the GraphSAGE calculation process, adjacent nodes may have similar embeddings when they share similar neighbors. To avoid this issue, in this paper we introduce skip connections to enhance the embedding process. As shown in Fig. 3, the output of each GraphSAGE layer is the concatenation of the current layer together with the outputs of all the previous layers. Assume $1 \leqslant k \leqslant K$ is the number of GraphSAGE layers ($K = 8$ in this paper), $\mathbf{H}_k^{'}$ is the output embedding matrix of the $k^{\text{th}}$ layer, $\mathbf{H}_k$ is the final embedding matrix after all $k$ layers. Then we have

$$\mathbf{H}_k = concat(\sum_{i=1}^{k-1} w_{i,k} H_i, H_k^{'}), \forall k = 1, 2, ..., K \tag{6}$$

where $w_{i,k}$ is a trainable weight on the skip connection from layer $i$ to layer $k$. Because $\mathbf{H}_k^{'}$ is essentially the embedding of the $k$-hop neighborhood, such skip connections can ensure the final embedding results include all the important subgraph features and avoid significant information losses throughout the deep learning model, which enhances the embedding performance.

## 2.4 Motif Searching

As mentioned in Section 2.2, the frequency of a given query graph $G_q$ in the target graph $G_T$ can be quickly approximated based on the ordered embedding space. However, as the number of nodes increases, the number of possible structures of $G_q$ increases exponentially, which is computationally expensive to calculate the frequency of every candidate to find out the motif. To address this issue, we introduce a greedy-search-based algorithm to iteratively grow the motif from a seed graph to improve the efficiency, as shown in Fig. 2(b). Detailed process is as follows.

(1) *Initialization*. We randomly select a node $G_0$ in $G_T$ as the seed graph and start to grow the motif.

(2) *Motif growing*. As illustrated in Fig. 5, starting from $G_0$, we iteratively include each of the adjacent nodes of $G_0$ as well as the edges connected to formulate the expanded graph $G_1$. All the candidates of $G_1$ will be embedded into the ordered embedding space by the trained encoder to calculate their frequency. The candidate with the highest frequency will be selected as the updated motif. This process continues until the number of nodes of the motif reaches the target, showing a monotinic "walking path" in the ordered embedding space from the lower left to the upper right, as shown in Fig. 2(b).

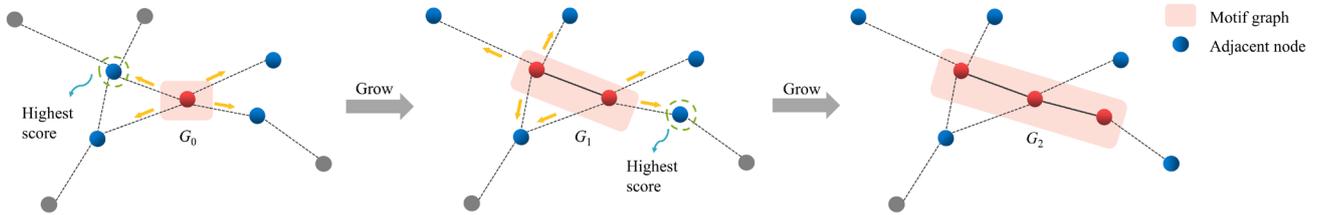

Fig. 5. Illustration of the proposed motif growing algorithm.

(3) *Repeative trials*. Because the initial node $G_0$ is randomly selected from the graph, different initialization may lead to different motif growing results. To offset the randomness, we repeat the above process (1) and (2) for multiple times and obtain a large number of candidate motifs with the given node size.

(4) *Motif finalization*. The candidate motifs obtained in step (3) are further compared regarding isomorphism and ranked in descending order based on their frequency to find the most frequent candidate as the final motif.

Note that in practice, the target graph database may include multiple graphs instead of a single one. Consequently, the motif needs to be searched across all target graphs. To this end, a discrete probability distribution that is proportional to the size of the target graphs will be created and assigned to each target graph. When establishing the reference vectors and performing the motif growing algorithm, the target graph will first be selected based on the discrete probability distribution, ensuring large graph will have a better chance to be searched.

## 3. Case Study

### 3.1 Test Case Setup

In this paper, we implement the proposed motif mining method on a power system database including 61 circuit models, as shown in Table I. We evaluate the effectiveness of the proposed method from two aspects. First, we examine the motif mining results to see if they are in consistence with the domain knowledge in power system planning. Second, we compare the motif mining results with a traversal searching algorithm, which can be considered as the ground truth, to see if the results are accurate. For each circuit model, we only keep the static topological information and eliminate irrelevant features such as branch resistance, reactance, thermal limit, etc. Note that the node type (PV, PQ, ect.) and voltage level are considered relevant to the power system topology and will be remained as nodal features. The algorithm is developed in PyTorch environment on a personal desktop with NVIDIA GeForce RTX 3090 GPU and Intel Xeon Gold 6330 CPU.

Table I Summary of the testing systems

| No. | System name | Description |
|---|---|---|
| 1 | IEEE-4, IEEE-13, IEEE-34, IEEE-37, IEEE-123 | Unbalanced radial feeders [28] |
| 2 | IEEE-14, IEEE-RTS-24, IEEE-30, IEEE-57, IEEE-118, IEEE-145, IEEE-300 | Transmission test systems [28] |
| 3 | case4_dist, case18, case22, case33bw, case69, case69, case141, | Small radial distribution systems [31] |
| 4 | case4gs, case5, case6ww, case9, case9Q, case9target, case30pwl, case30Q, case39 | Small transmission systems [29] |
| 5 | J1, K1, M1, Ckt5, Ckt7, Ckt24 | Real distribution system feeders [30] |
| 6 | case1888rte, case1951rte, case2848rte, case2868rte, case6468rte, case6470rte, case6495rte, case6515rte | Real French power systems [31] |
| 7 | case89pegase, case2869pegase, case9241pegase, case13659pegase | European high-voltage transmission networks [31][32] |
| 8 | ACTIVSg200, ACTIVSg500, ACTIVSg2000, ACTIVSg10k, ACTIVSg25k, ACTIVSg70k, SyntheticUSA | ACTIV synthetic grids [33] |
| 9 | case2383wp, case2736sp, case2737sop, case2746wop, case2746wp, case3012wp, case3120sp, case3375wp | Poland power systems [31] |

### 3.2 Construction and Analysis of the Ordered Embedding Space

The power system models are first decomposed into 10,000 small neighborhoods with the number of nodes ranging from 3 to 25. These neighborhoods are used to train the encoder network in Section 2.3 guided by the loss function in (2). These neighborhoods are encoded by the trained encoder to formulate the reference vectors in the ordered embedding space. Here we randomly select 1000 reference vectors as a representative and plot out their 2D distribution after dimension reduction using Principal Component Analysis (PCA), as shown in

Fig. 6. Note that static nodal features such as voltage level, node type are considered as part of the topology.

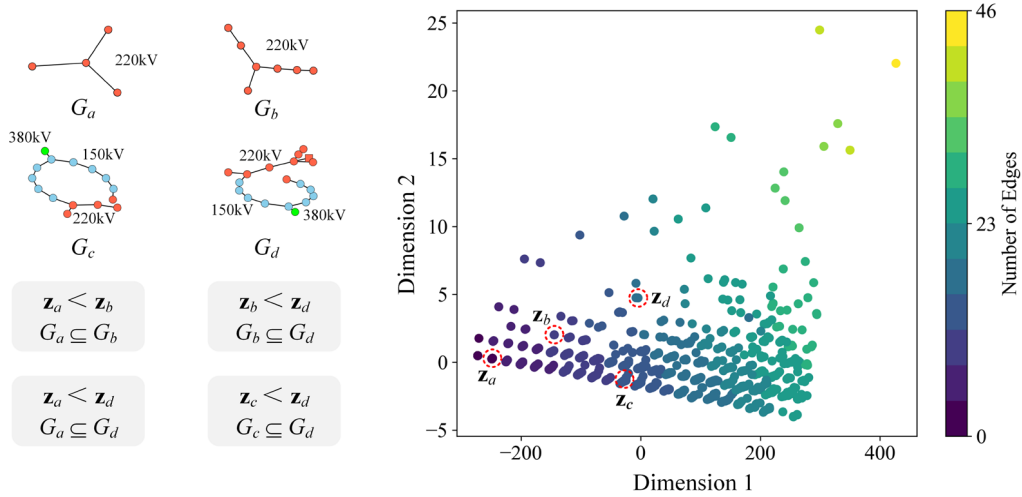

Fig. 6. 2D demonstration of 1000 reference vectors in the ordered embedding space after dimension reduction using PCA. $G_a$, $G_b$, $G_c$, $G_d$ are 4 selected neighborhoods, while $\mathbf{z}_a$, $\mathbf{z}_b$, $\mathbf{z}_c$, $\mathbf{z}_d$ are their corresponding vector embeddings. Circles in the topologies represent PQ nodes, squares represent PV nodes, and different colors represent different voltage levels.

From Fig. 6, we can see that small graphs (dark color dots) are generally positioned at the lower left of large graphs (light color dots). This is because only small graphs could be the subgraph of large graphs, but not vice versa. For example, $\mathbf{z}_a$ is on the lower left of $\mathbf{z}_b$, indicating $G_a$ is a subgraph of $G_b$ which can be validated by comparing their topologies. Similarly, $\mathbf{z}_a$, $\mathbf{z}_b$ and $\mathbf{z}_c$ are all on the lower left of $\mathbf{z}_d$, so that $G_a$, $G_b$ and $G_c$ are all subgraphs of $G_d$. Meanwhile, we also observe that there are many large graphs locating at the lower right of small graphs, such as the relative location between $\mathbf{z}_a$ and $\mathbf{z}_c$. This means there is no subgraph isomorphic relationship between $G_a$ and $G_c$. This result indicates the encoder is well-trained and the embedding space is highly ordered.

We further calculate the norms of the embedding vectors and demonstrate their changes along with the graph size, as shown in Fig. 7. From Fig. 7(a), we can see that the embedding norm increases linearly with the number of graph nodes, demonstrating the embedding space is highly ordered. Violations exists especially for large graphs with more nodes, because large graphs usually have complex and diversified structures that have no subgraph isomorphic relationship with other large graphs. Similar conclusion can be obtained in Fig. 7(b) with more violations observed when indexed by the number of graph edges. In Fig. 7(c), we plot out the heat map for the embedding norms dependent on both the number of nodes and edges to show the comprehensive pattern. We can see that the largest embeddings usually happen when the number of graph nodes and edges are close. This means radial structures with few loops have the best chance to be subgraph isomorphic with other graphs, which is in consistent with the domain knowledge that most power system fractions are radial.

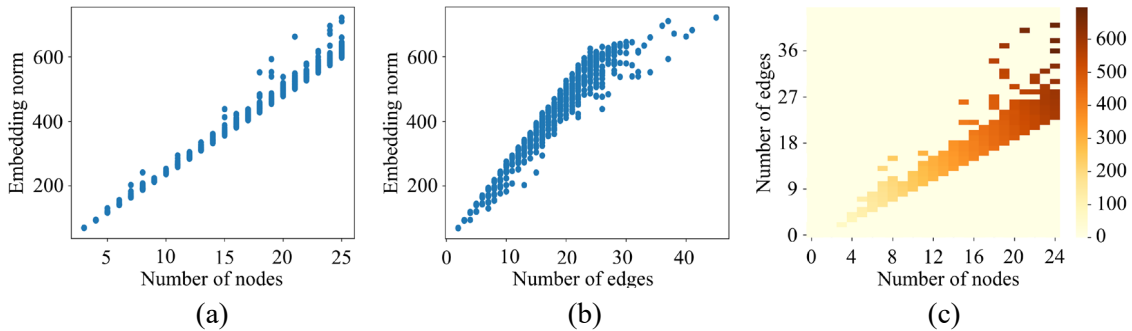

Fig. 7. The relationship between graph size and the embedding norm. Embedding norm changes along with (a) the number of graph nodes, (b) the number of graph edges, and (c) both the number of graph nodes and edges.

### 3.3 Motif Mining Results

After the ordered embedding space is obtained, we perform the motif searching algorithm in section 2.4 to find the most frequent topological structures across the whole database in Table I. In this paper we search motifs with 3-10 nodes. Results of the most frequent motifs are shown in Fig. 8.

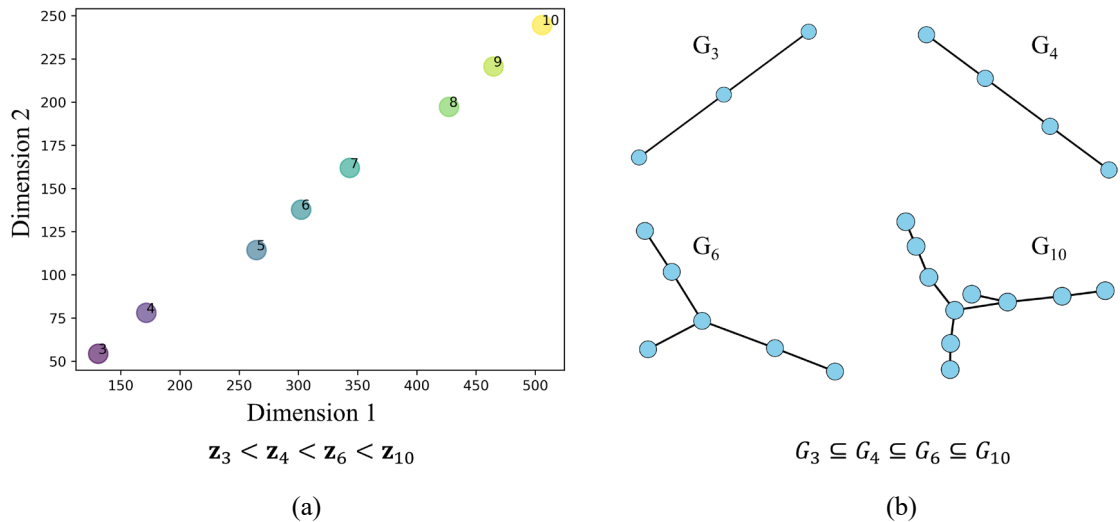

|  (a)  |  (b)  |

Fig. 8. Motif mining reuslts. (a) Vector embeddings of the most frequent motifs with 3-10 nodes. (b) Motifs with 3, 4, 6 and 10 nodes.

From Fig. 8(a), we can see that the vector embeddings of the obtained motifs strictly follow the "upper left" rule, indicating there are subgraph isomorphic relationships among the motifs. For instance, the motifs with 3, 4, 6, 10 nodes are shown in Fig. 8(b). This is reasonable because large graphs that derived from frequent small graphs have higher chance to become motifs.

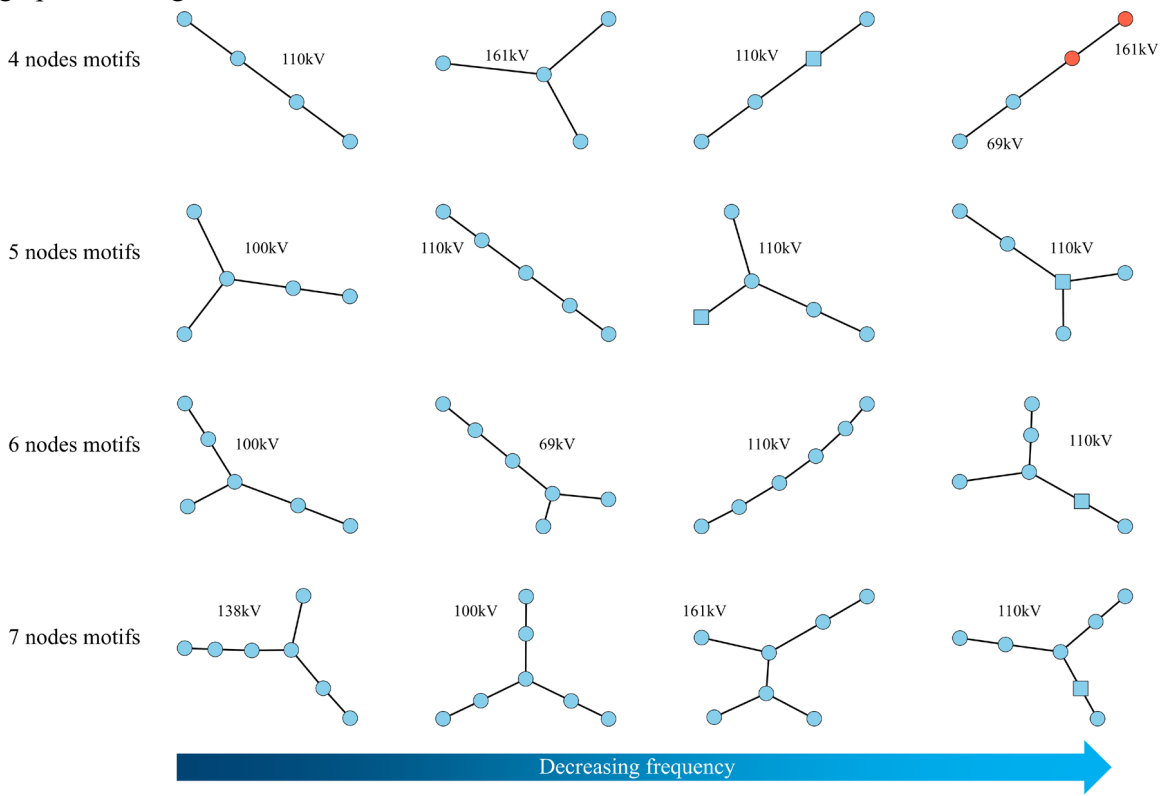

Fig. 9. Top-4 motifs with 4, 5, 6, 7 nodes with decreasing frequency.

We further plot out the top-4 frequent motifs with different node numbers in decreasing order by frequency, shown as Fig. 9. We have the following observations:

- Linear and radial structures are the most common microstructures in power system topology. As the number of nodes increases, radial structures with multiple branches becomes the mainstream. This result is in consistent with the power system designing principles because branches are needed to increase the power supply region and enhance the system reliability.

- Most of the motifs are in the same voltage level. Because of the hierarchical structure of power systems, most of the test cases in Table I focus on either transmission or distribution with only a few voltage level included, instead of modeling the whole system. Particularly, 110kV is the most frequent voltage level of motifs across the whole database.

### 3.4 Distribution System Analysis: Rural and Urban

We implement the proposed method to find motifs for rural and urban power distribution feeders respectively and compare their differences. The rural feeder is a detailed feeder model designed for studying neutral-to-earth voltage effect [34]. The main voltage of the feeder is 13.2kV supplying residential loads. For the urban feeder model, we select the IEEE-342 test feeder which simulates a moderate urban power system [35]. Here we search motifs for the two test feeders within the range of 3-30 nodes. 8 obtained motifs are shown in Fig. 10 as an example.

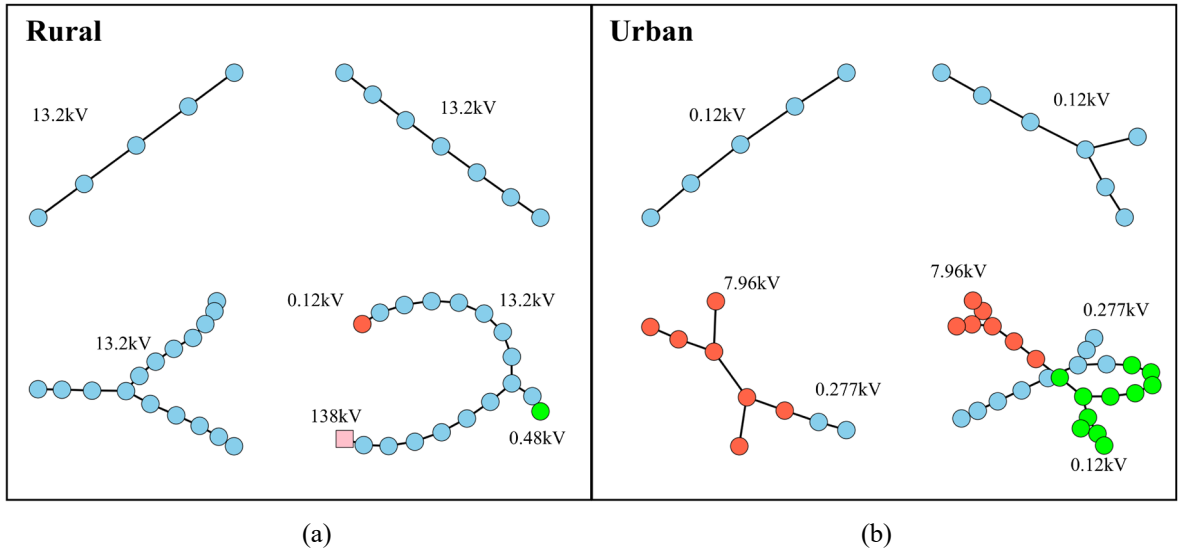

(a)            (b)

Fig. 10. Samples of the motif mining results. (a) Rural feeder. (b) Urban feeder.

We can see distinct characteristics of the two feeders from Fig. 10. The rural feeder is a typical radial distribution system that has long distance but limited number of branches to supply the sparse residential loads. On the contrary, the urban feeder involves in more voltage levels to fit for different types of loads, and has more complex structures including loops to guarantee the power supply reliability. Such an observation is consistent with the domain knowledge of power system planning.

To further evaluate the accuracy of the proposed method, we introduce the VF2 algorithm in [36], which is a traversal-searching-based graph matching algorithm, to exhaustively count the frequency of the obtained motifs as the ground truth. Results for both rural and urban motifs with node 3-6 are shown in Table II as an example. we can see that the rankings of the top-3 motifs obtained by the proposed method is generally consistent with the traversal searching results, especially for the most frequent motifs, demonstrating the accuracy of the proposed AI-powered motif mining method.

Table II Validation of the motif mining results

| | Nodes | Proposed method (ranking) | VF2 results (frequency) | Nodes | Proposed method (ranking) | VF2 results (frequency) |
|---|---|---|---|---|---|---|
| **Rural** | 3 | 1 | 152 | 4 | 1 | 146 |
| | | 2 | 4 | | 2 | 6 |
| | | 3 | 5 | | 3 | 1 |
| | 5 | 1 | 150 | 6 | 1 | 138 |
| | | 2 | 5 | | 2 | 4 |
| | | 3 | 2 | | 20 | 3 |
| **Urban** | 3 | 1 | 574 | 4 | 1 | 682 |
| | | 2 | 388 | | 2 | 56 |
| | | 3 | 56 | | 3 | 32 |
| | 5 | 1 | 858 | 6 | 1 | 1070 |
| | | 2 | 112 | | 2 | 168 |
| | | 3 | 132 | | 3 | 152 |

### 3.5 Transmission System Analysis: USA and Europe

We implement the proposed motif mining method to analyze the topological characteristics of transmission systems in USA and Europe, respectively. We use the case ACTIVSg70k in [33]to represent the USA transmission system, which is a large-scale synthetic circuit model with 70,000 buses built on eastern United States. As a comparison, we use the case 13659pegase in [31] as a representative for the Europe transmission system, which is a pan-European synthetic circuit model with 13,659 nodes. Samples of the motif mining results for the two test cases are shown in Fig. 11. We can see distinct characteristics with respect to voltage level and topology.

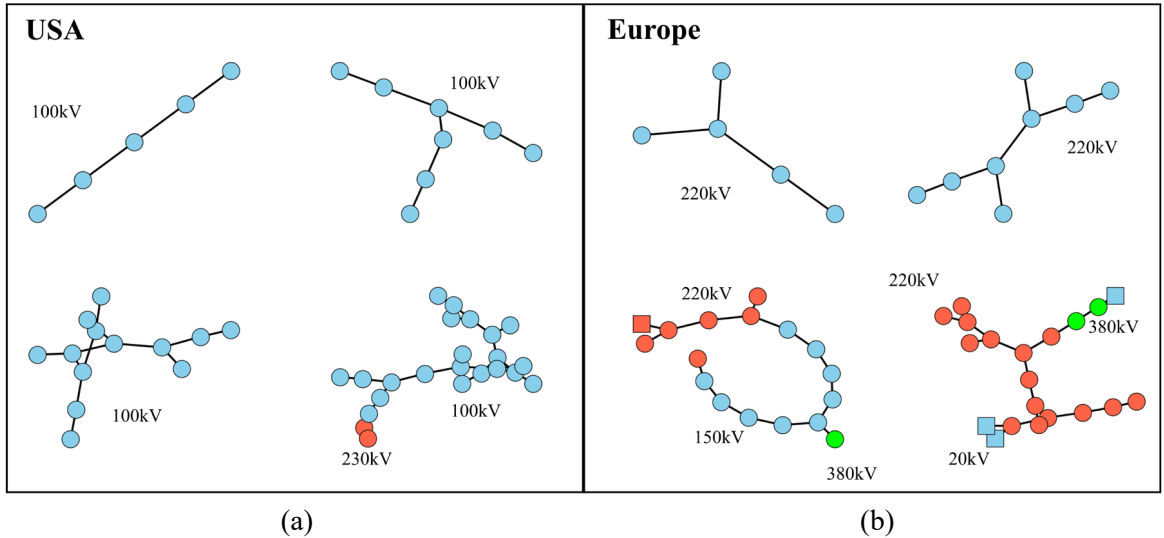

(a)                                                                 (b)

Fig. 11. Examples of the motif mining results. (a) USA transmission system. (b) Europe transmission system.

Similar to Fig. 7(c), we plot out the heat maps of the embedding norms of the obtained motif mining results for both the USA and Europe systems, with respect to the number of nodes and edges, as shown in Fig. 12. We can see that the motifs of the USA transmission system have comparable number of nodes and edges, indicating that the major topology structures are linear or radial. This is common especially in the vast rural areas in central USA. On the contrary, the number of edges of the motifs in the Europe transmission system are much larger than that in USA, indicating more complex and diversified topology structures.

Note that the motif searching process is essentially an NP-hard problem. For large graphs such as the transmission systems in this section, the traversal-searching-based algorithms such as VF2 will lead to exponentially-increasing computation costs and are no longer applicable in practice. in contrast, the computation efficiency of the proposed AI-powered motif searching algorithm is independent with the graph size, making it particularly suitable for solving large graph problems.

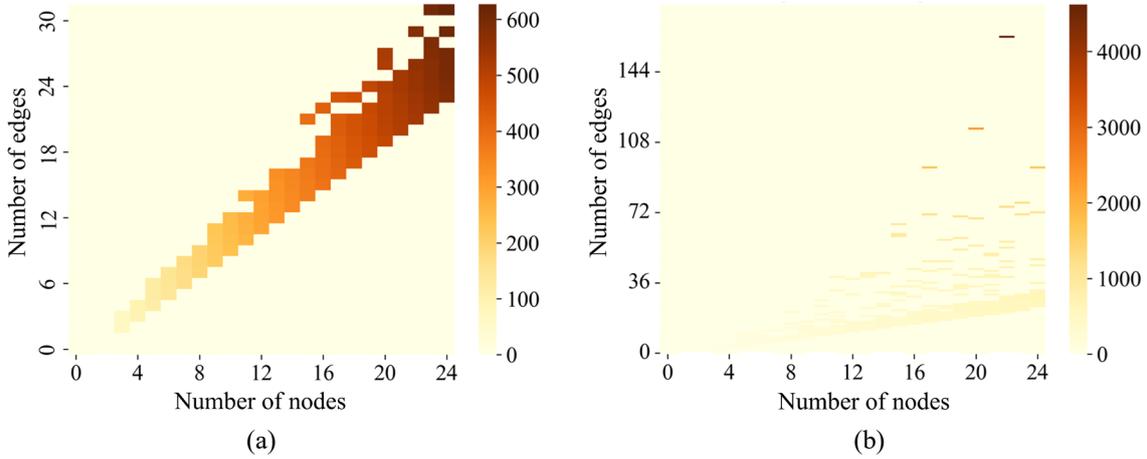

(a)                                                  (b)

Fig. 12. The heat maps of averaged embedding norms under different nodes and edges. (a) USA. (b) Europe.

## 4. Conclusions

In this paper, a two-stage AI-powered motif mining method is introduced to enable efficient and wide-range motif analysis for large-scale power systems. A representation learning model based on graphSAGE network is designed to achieve ordered embedding for the power system topology, based on which the subgraph matching problem can be converted into a vector comparing problem to improve the motif mining efficiency. Then a greedy-search-based algorithm is proposed to iteratively grow motifs from a ramdom initialization node in the target graph, so that the traditional time-consuming traversal searching process can be avoided. Case study is based on a power system database composed by 61 circuit models. The effectiveness of the proposed method is demonstrated from two aspects. First, the motif mining results for real-world transmission and distribution systems are analyzed and are found to be in consistence with the domain knowledge in power system planning. Second, the motif mining results are in consistence with the VF2 results, which is a traversal searching algorithm and can serve as the ground truth. The proposed method is efficient and interpretable, bringing new opportunities for the power system topological analysis and serving as a reference for the application of interpretable machine learning in power systems.

Future work may focus on implementing the proposed motif mining method to solve downstream tasks to demonstrate its engineering values, such as power system vulnerability analysis and synthetic circuit model generation.

## 5. Acknowledgement

This work was supported in part by National Natural Science Foundation of China under Grant 52307121, and in part by Shanghai Sailing Program under Grant 23YF1419000.